\documentstyle[aps,prd,epsf]{revtex}
\newcommand{\be}{\begin{equation}}
\newcommand{\ee}{\end{equation}}
\newcommand{\ben}{\begin{eqnarray}
\displaystyle}
\newcommand{\een}{\end{eqnarray}}

\newcommand{\cD}{{\cal D}}
\newcommand{\cM}{{\cal M}}
\newcommand{\cJ}{{\cal J}}

\newcommand{\p}{\partial}
\newcommand{\na}{\nabla}

\newcommand{\tiA}{{\tilde A}}
\newcommand{\tiB}{{\tilde B}}
\newcommand{\tiG}{{\tilde G}}

\newcommand{\ep}{\epsilon}
\newcommand{\bep}{\bar \epsilon}

\newcommand{\om}{\omega}
\newcommand{\ga}{\gamma}

\begin{document}

\title{Uniqueness Theorem for Static Dilaton $U(1)^2$ Black Holes}

\author{Marek Rogatko}

\address{Technical University of Lublin \protect \\
20-618 Lublin, Nadbystrzycka 40, Poland \protect \\
rogat@tytan.umcs.lublin.pl \protect \\
rogat@akropolis.pol.lublin.pl}

\date{\today}

\maketitle
%%%%%%%%%%%%%%%%%%%%%%%%%%%%%%%%%%%%%%%%%%%%%%%%%%%%%%%%%%%%%%%%%%

\begin{abstract}
Using a sigma model formulation of the field equations as
on a two-dimensional manifold we provide the proof of
a black hole uniqueness solutions in
$N=4, d=4$ supergravity subject to certain boundary conditions.
We considered the black hole solutions both in 
$SU(4)$ and $SO(4)$ versions of the
underlying theory.
\end{abstract}

%%%%%%%%%%%%%%%%%%%%%%%%%%%%%%%%%%%%%%%%%%%%%%%%%%%%%%%%%%%%%%%%%%%%

\baselineskip=18pt
\section{Introduction}
During the last years there has been a considerable resurgance of
mathematical studies of topics related to the black hole equilibrium
states. Not only restricted to the pure vacuum Einstein or
Einstein-Maxwell theory but also including nonlinear matter models,
general sigma models or fields occuring in the low energy limits of
the superstring theories.

In his pioneering investigations  \cite{iso,is1} Israel established 
the uniqueness of the Schwarzschild metric and its Reissner-Nordstr\"om
generalization
as static asymptotically flat solutions of the Einstein and Einstein-Maxwell
vacuum field equations subject to the condition of regularity outside a well
behaved {\it ergosurface} where the static Killing vector becomes null.
In view of the subsequent demonstration \cite{caj,ca0} that in the 
static case
(though not more generally) the {\it ergosurface} will be an event
horizon, it follows that
the Schwarzschild and Reissner-Nordstr\"om solutions are the only Einstein or 
Einstein-Maxwell (non-extreme) solutions that satisfy the condition of being 
static black hole metrics in the strict modern sense of the term.
\par
Israel's article \cite{iso} 
was the foundation of the next works of M\"uller zum Hagen {\it et al.}
\cite{mi} and Robinson
\cite{ro} establishing the generalization of Israel's theorem of
the uniqueness of the Schwarzschild black hole solution.
\par
The uniqueness results for rotating configurations, i.e., for stationary and 
axisymmetric black hole spacetimes were achieved by Carter in
\cite{cak} and completed by Hawking and Ellis
 \cite{haw} and the next works of Carter \cite{ca0,ca},
Robinson \cite{rok} and Wald \cite{wak}.
These works were devoted to the vacuum case.
Bunting \cite{bun} and Mazur \cite{ma} generalize these results to the 
electromagnetic case.
Bunting's approach was based on using a general class of harmonic mappinng
between Riemannian manifolds. In Mazur's proof the key point was the
observation that the Ernst Eqs. described a nonlinear sigma model on a
symmetric space.
The review of the new methods presented by Bunting and Mazur
was discussed in \cite{car}.
\par
Bunting and Masood-ul-Alam \cite{bu} used
the positive mass theorem \cite{ya,wi} to prove the 
uniqueness, the spherical symmetry
or the nonexistence
for several static black hole solutions of the Einstein's Eqs. 
In Ref.\cite{bu} the afore mentioned technique
was explored to prove the nonexistence
of multiple black holes in the asymptotically Euclidean 
static vacuum spacetime.
They found the conformal transformation which caused that the mass of 
the spatial part of the metric was equal to zero, but the scalar curvature
tensor was non-negative.
Ruback \cite{ru} applied this technique to prove the uniqueness theorem for
charged black holes in static Einstein-Maxwell spacetime.
Masood-ul-Alam \cite{ma1} gave an alternative, much simpler, a rigorous proof
that the unique non-degenerate electrovac static black hole metrics are the  
Reissner-Nordstr\"om  family. It was done without assuming the connectedness
of the event horizon.
The further generalization of the uniqueness proof for the static electrovac
black holes including the case of a non-vanishing magnetic charge was proposed
by Heusler \cite{he1}. He used
a generalization of the conformal factor
and established the nonexistence of multiple black hole solutions of
Einstein-Maxwell system with electric and magnetic fields in a static,
asymptotically flat spacetime.
Heusler \cite{he2} demonstrated also
the uniqueness of multiple black hole solutions for any self-coupled, 
stationary scalar mapping (sigma-model) with nonrotating horizon.
\par
Using the fact that Einstein-Abelian gauge field Eqs. can be formulated
as a sigma model on the adequate K\"ahler manifold, G\"urses \cite{gu} found
that $(n-1)$ Abelian gauge charged Kerr black hole is a unique
stationary black hole solution of Einstein-Abelian gauge field Eqs.
In Ref.\cite{gu1}, he also proved that, the boundary value problems of some
sigma models in a non-Riemannian background have unique solutions.
\par
Quite recently a uniqueness theorem was extended to the case of the
Ernst solution and C-metric \cite{we}.
\par
Uniqueness theorems for black holes are closely related to the
problem of staticity. Lichnerowicz \cite{li} was the first who considered the
idea of staticity for a stationary perfect fluid, locally static in the
sence that its flow vector was connected with the Killing vector. The
next extensions was atributed to Hawking \cite{haw} (vacuum case) and the
Hawking's {\it strong rigidity theorem} 
\cite{hel}
emphasized that the event
horizon of a stationary black hole had to be a Killing horizon.
Recently, Sudarsky and Wald \cite{wa}, by means of the notion of an
asymptotically flat maximal slice with a compact interior, obtained the
conditions that the solution of Einstein-Yang-Mills Eqs. is static when
it had a vanishing Yang-Mills electric field on static hypersurfaces.
They also reached to the conclusion that nonrotating Einstein-Maxwell
black hole must be static when it has a vanishing magnetic field on
static slices \cite{wa1}.
\par
For a recent review concerning various aspects of uniqueness
theorems for nonrotating and rotating black holes see \cite{book}, while
the mathematical rigor of the afore mentioned problems
has been studied in the review articles provided by 
Chru\'sciel \cite{chr1,chr2}.

%%%%%%%%%%%%%%%%%%%%%%%%%%%%%
\medskip
\noindent
Recently, there has been an active period for constructing black hole
solutions in the string theory which seems the most promising
for a theory of quantum gravity (see \cite{hu}
for a recent review of the subject).
\par
The uniqueness of static, charged dilaton black hole solutions
in the low energy
string theory was certified by Masood-ul-Alam \cite{ma2}.
He found a conformal spatial metric which had the sufficient 
properties for existing
suitable Dirac spinors. By means of the Lichnerowicz Eq. it was 
shown
that these spinors are constant.
\par
The alternative proof of a uniqueness of a static charged dilaton
black hole was provided by G\"urses and Sermutlu \cite{gs}. They used
a sigma model formulation of equations of motion.
The problem of black hole solutions and their uniqueness
in axion-dilaton gravity
was
studied by Bowick {\it et al.} \cite{bow}. They managed to find
uniqueness theorem in the case of the minimal coupling of axion field
to gravity. Cambell {\it et al.} \cite{cam} showed the existence of
axion {\it hair} for a Kerr black hole and calculated it explicitily
in the case of a slow motion. They considered axion fields
with a Lorentz Chern-Simons coupling to gravity.
A dilaton coupling to axion fields strengths were considered in
\cite{mig}, where the authors calculated dilaton {\it hair}
arising from the specific axion source.
\par
Wells \cite{we1} wrote down the analogue of the Robinson's identity for
dilatonic black holes which allowed him to prove uniqueness theorem
for a class of accelerating stringy black holes.
\par
The problem of staticity theorems in Einstein-Maxwell axion-dilaton
gravity was studied by Rogatko in Ref.\cite{r1}, where the modified Carter
arguments were used to find staticity conditions for fields and the
metric. It was found, in Ref.\cite{r2}, that static black
hole solutions in the above theory had vanishing electric and
axion-electric fields on static slices.
\par
Our paper is organized as follows. In Sec.II we present the equations of
motion for the bosonic sector of $N = 4, d= 4$ supergravity in a static
axially symmetric spacetime. Introducing the adequate forms of the
pseudopotentials and complex scalars enables us to write Eqs. of motion
as two complex Ernst-like Eqs. Then, using the matrix formulation of
Ernst Eqs., conceived by G\"urses and Xanthopoulos \cite{gx}, we reached the
conclusion that two metrics satisfying the Eqs. of motion and having
the same boundary conditions must be equal to each other in all points
of the region of the two-dimensional
manifold. Which implies in turns, that all black hole solutions in
the theory under consideration subject to the same boundary conditions
are the same everywhere in the spacetime. We considered the $SU(4)$ and
$SO(4)$ versions of the underlying theory.
We conclude in Sec.III with a brief summary of our researches and their
implications.
%%%%%%%%%%%%%%%%%%%%%%%%%%%%%%%%%%%%%%%%%%%%%%%%%%%%%%%%%%%%%%

\section{Uniqueness of Black Hole Solutions}
\subsection{Doubly Charged Black Holes in $SU(4)$ version of $N = 4, d
= 4$ supergravity}
Superstring theories
provide interesting generalizations of the Einstein-Maxwell theory in
the so-called low energy limit. A dimensionally reduced superstring
theory can be described in terms of
$N=4, d = 4$ supergravity. It turned out, that one can
refer to the $SO(4)$ version \cite{so} or $SU(4)$ one \cite{su}.
In our paper, we shall consider 
bosonic sectors of these theories, taking into account 
two $U(1)$ gauge fields 
and a dilaton field $\phi$, called $U(1)^2$ theories.
We begin with the $SU(4)$ version of $N = 4, d = 4$ supergravity, which
the action is of the form \cite{so,kal}
\be
I_{SU(4)} = \int d^4 x \sqrt{-g} 
\left [ 
	R - 2(\na \phi)^{2}
- e^{-2\phi} 
\left (
F_{\alpha \beta} F^{\alpha \beta} +
G_{\ga \delta} G^{\ga \delta} 
\right )
\right ],
\label{act}
\ee
where the strengths of the gauge fields are descibed by
$F_{\mu \nu} = 2\na_{[\mu} A_{\nu]}$ and 
$G_{\alpha \beta} = 2\na_{[\alpha} B_{\beta] }$.
The resulting equations of motion, derived from the variational
principle, are as follows:
\ben
R_{\mu \nu} = e^{-2 \phi} \left (
2 F_{\mu \rho} F_{\nu}{}{}^{\rho} - {1 \over 2} g_{\mu \nu}F^2 \right ) +
e^{-2 \phi} \left (
2 G_{\mu \rho} G_{\nu}{}{}^{\rho} - {1 \over 2} g_{\mu \nu}G^2 \right ) +
2 \na_{\mu} \phi \na_{\nu} \phi, \\
\na_{\mu} \na^{\mu} \phi + {1 \over 2} e^{-2\phi} F^2 + 
{1 \over 2} e^{-2\phi} G^2 = 0, \\
\na_{\mu} \left ( e^{-2 \phi} F^{\mu \nu} \right ) = 0, \\
\na_{\mu} \left ( e^{-2 \phi} G^{\mu \nu} \right ) = 0.
\een
The black hole solutions in the theory under consideration were widely
discussed by Kallosh {\it et al.} in Ref.\cite{kal}. 
\par
Our main task will be to prove the uniqueness of
the obtained results. We want to provide some continuity with the
researches of G\"urses \cite{gu1,gs}, in some respects to generalize them,
we shall
present our analysis of the problem in a form and notation similar to
theirs. 
First, we shall formulate the corresponding Eqs. of motion as a
two-dimensional sigma model and prove the uniqueness of the static
solution under the same boundary conditions.
In order to do so we introduce the static axially symmetric line element
expressed as
\be
ds^2 = - e^{2 \psi} dt^2 + e^{-2 \psi} \left [
e^{2 \ga} \left ( dr^2 + dz^2 \right ) + r^2 d\phi^2 \right ],
\label{gij}
\ee
where $\psi $ and $\ga$ depended only on $r$ and $z$ coordinates.
The components of the $U(1)$ gauge strength tensors will be in the
direction of time $F_{\mu \nu} = (A_{0}, 0, 0, 0)$ and in the azimuthal
angle $G_{\alpha \beta} = (0, 0, 0, B_{\phi})$. In our further
considerations we assume that the components of the $U(1)$ gauge fields
are functions of $r$ and $ z$. Then,
the resulting Eqs. of motion are as follows:
\be
\na^2 \phi - e^{-2\psi - 2\phi} \left (
A_{0, r}^2 + A_{0, z}^2 \right ) + {e^{2\psi - 2 \phi} \over r^2}
\left ( B_{\phi, r}^2 + B_{\phi, z}^2 \right ) = 0,
\label{em1}
\ee
\be
\na^2 \psi - e^{-2\psi - 2\phi} \left (
A_{0, r}^2 + A_{0, z}^2 \right ) - {e^{2\psi - 2 \phi} \over r^2}
\left ( B_{\phi, r}^2 + B_{\phi, z}^2 \right ) = 0,
\label{em2}
\ee
\be
\na^2 A_{0} - 2 \left ( \psi_{, r} + \phi_{, r} \right ) A_{0, r}
- 2 \left ( \psi_{, z} + \phi_{, z} \right ) A_{0, z} = 0,
\label{em3}
\ee
\be
\triangle B_{\phi} +
2 \left ( \psi_{, r} - \phi_{, r} \right ) B_{\phi, r} +
2 \left ( \psi_{, z} - \phi_{, z} \right ) B_{\phi, z} = 0,
\label{em4}
\ee
\be
e^{-2\psi - 2\phi} \left (
A_{0, r}^2 - A_{0, z}^2 \right ) + {1 \over r^2} e^{2\psi - 2 \phi} 
\left ( B_{\phi, r}^2 - B_{\phi, z}^2 \right ) +
\left ( \phi_{,z}^2 - \phi_{, r}^2 \right ) =
\psi_{, r}^2 - \psi_{, z}^2 - {\ga_{, r} \over r},
\label{em5}
\ee
\be
{\ga_{, z} \over r} - 2 \psi_{, r} \psi_{, z} =
- 2 e^{-2\psi - 2\phi} A_{0, r} A_{0, z} + {2 \over r^2} e^{2\psi - 2 \phi} 
B_{\phi, r} B_{\phi, z} + 2 \phi_{, r} \phi_{, z},
\label{em6}
\ee
where $\na^2$ is the Laplacian operator in the $(r, z)$ coordinates,
namely, $\na^2 = \p_{r}^2 + \p_{z}^2 + {1 \over r} \p_{r}$
and $\triangle = \p_{r}^2 + \p_{z}^2 - {1 \over r}\p_{r}$.\\
From Eq.(\ref{em5}) or (\ref{em6}) one can determine the function $\ga$
if $\psi, \phi, A_{0}, B_{\phi}$ are known. Hence, the essential
part of the Eqs. of motion consists of Eqs.(\ref{em1}-\ref{em4}).
\par
Now, let us define the quantities
\be 
E = - \phi - \psi, \qquad M = \psi - \phi,
\ee
and
\be
\tiA_{0} = {A_{0} \over \sqrt{2}}, \qquad 
\tiB_{\phi} = {i B_{\phi} \over \sqrt{2}}.
\ee
Consistently with the above definitions,
Eqs.(\ref{em1}-\ref{em4}) can be rewritten in the forms
\be
\na^2 E + e^{2 E} \na \tiA_{0} \na \tiA_{0} = 0,
\label{a0}
\ee
\be
\na^2 \tiA_{0} + 2 \na E  \na \tiA_{0} = 0,
\label{a1}
\ee
\be
\na^2 M + e^{2 M} \na \tiB_{\phi}  \na \tiB_{\phi} = 0,
\label{a2}
\ee
\be
\triangle \tiB_{\phi} + 2 \na M \na \tiB_{\phi} = 0.
\label{a3}
\ee
Thus, we obtain two pairs of Eqs., one described in terms of 
$\tiA_{0}, E$ and the other for $\tiB_{\phi}$ and $M$.\\
We observe that Eq.(\ref{a1}) allows us to define a pseudopotential
$\om_{(A)}$, given by
\be
\om_{(A) r} = r e^{2 E} \tiA_{0, z}, \qquad
\om_{(A) z} = - r e^{2 E} \tiA_{0, r},
\label{wa}
\ee
while from Eq.(\ref{a3}), one can establish the following
$\om_{(B)}$ pseudopotential:
\be
\om_{(B) r} =  { - e^{2 M} \tiB_{\phi, z} \over r}, \qquad
\om_{(B) z} = {e^{2 M} \tiB_{\phi, r} \over r}.
\label{wb}
\ee
Then,
we want to rewrite the Eqs.(\ref{a0}-\ref{a3}) 
in the forms similar to the Ernst ones.
In order
to do this, we introduce two complex scalars, determined by
\ben
\label{e1}
\ep_{1} &=& r e^E + i \om_{(A)}, \\
\ep_{2} &=&  e^{M} + i \om_{(B)}.
\een
Now, Eqs.(\ref{a0}-\ref{a1}) and (\ref{a2}-\ref{a3}) can be 
arranged into the following
two complex Eqs.:
\ben
\label{ee1}
\left ( \bep_{1} + \ep_{1} \right ) \na^2 \ep_{1} =
2 \na \ep_{1} \na \ep_{1}, \\
\left ( \bep_{2} + \ep_{2} \right ) \na^2 \ep_{2} =
2 \na \ep_{2} \na \ep_{2},
\label{ee2}
\een
where a bar denotes complex conjugation.\\
Eqs.(\ref{ee1}) and (\ref{ee2}) are two Ernst Eqs., which each of them
combine in a convenient and a symmetric fashion the two Eqs. governing
$\phi, \psi$ and the adequate gauge field.
Each of them defined a sigma model
on $SU(2)/U(1)$.\\
It was shown in Ref.\cite{gx},
that the various combinations of Ernst's Eqs. were
included in the single matrix Eq. In our case, the matrix Eq. is
determined by
\be
\p_{r} \left [ P_{(i)}^{-1} \p_{r} P_{(i)} \right ]
+ \p_{z} \left [ P_{(i)}^{-1} \p_{z} P_{(i)} \right ] = 0,
\label{ma}
\ee
where the subscript $(i)$ in $P$ matrix 
refers respectively to $\tiA, \tiB$ gauge
fields. 
The explicit form  of the matrices are given by
\be
P_{(A)} = {1 \over r e^E} \pmatrix{1 & \om_{(A)} \cr \om_{(A)} &
r^2 e^{2 E} + \om_{(A)}^2 \cr},
\qquad
P_{(B)} = {1 \over e^M} \pmatrix{1 & \om_{(B)} \cr \om_{(B)} &
e^{2 M} + \om_{(B)}^2 \cr}.
\ee
One can check that, the above matrix Eqs. when written
out explicitly in terms of its elements constitute four Eqs., all of
them are various combinations of Eqs. (\ref{ee1}-\ref{ee2}) and Eqs.
the complex conjugate of them. 

In order to prove a uniqueness theorem we shall follow the line
described by G\"urses \cite{gu1}. 
To begin with, one should assume enough
differentiability for the matrices components in a region $\cD$
of the two-dimensional manifold $\cM$ with boundary $\p \cD$.
Let $P_{(i) 1}$ and $P_{(i) 2}$ will be two different solutions of Eqs.
(\ref{ma}) respectively for the cases of $\tiA$ and $\tiB$ 
gauge fields, than the difference
of their Eqs. will have be given by
\be
\na \left (  P_{(i) 1}^{-1} \left ( \na Q_{(i)} \right )
P_{(i) 2} \right ) = 0, 
\label{dif}
\ee
where $Q_{(i)} = P_{(i) 1} P_{(i) 2}^{-1}$.
Multiplying Eq.(\ref{dif}) by $Q_{(i)}^{\dagger}$, one arives at the
expression
\be
\na^2 q_{(i)} = Tr \left [
\left ( \na Q_{(i)}^{\dagger} \right ) P_{(i) 1}^{-1}
\left ( \na Q_{(i)} \right ) P_{(i) 2} \right ],
\label{g}
\ee
where $q = Tr Q$. Taking into account hermicity and
positive definiteness of the matrices $P_{(i) 1}$ and $P_{(i) 2}$,
we can postulate the form of the above matrices, satisfying
\be
P_{(A) \alpha} = A_{\alpha} A_{\alpha}^{\dagger},
\qquad
P_{(B) \alpha} = B_{\alpha} B_{\alpha}^{\dagger},
\label{pa}
\ee
where $\alpha = 1, 2$. The explicit forms of the matrices
$A_{\alpha}$ and $B_{\alpha}$ can be established as follows:
\be
A_{\alpha} = {1 \over \sqrt{r} e^{E_{\alpha}/2}} \pmatrix{
1 & 0 \cr \om_{(A) \alpha} & r e^{E_{\alpha}} \cr},
\qquad
B_{\alpha} = {1 \over e^{M_{\alpha}/2}} \pmatrix{
1 & 0 \cr \om_{(B) \alpha} & e^{M_{\alpha}} \cr}.
\label{aa}
\ee
Using relation (\ref{pa}) one can rewrite Eq.(\ref{g}) in the form
\be
\na^2 q_{(i)} = Tr \left ( \cJ_{(i)}^{\dagger} \cJ_{(i)} \right ),
\label{jj}
\ee
where $ \cJ_{(A)} = A_{1}^{-1} (\na Q_{(A)}) A_{2}$ and
$\cJ_{(B)} = B_{1}^{-1} (\na Q_{(B)}) B_{2}$.
Thus, the explicit versions of $q_{(i)}$ are given by
\ben
q_{(A)} &=& 2 + {1 \over r^2 e^{E_{1} + E_{2}}}
\left [ \left ( \om_{(A)1} - \om_{(A)2} \right )^2 +
r^2 \left ( e^{E_{1}} - e^{E_{2}} \right )^2 \right ], \\
q_{(B)} &=& 2 + {1 \over e^{M_{1} + M_{2}}}
\left [ \left ( \om_{(B)1} - \om_{(B)2} \right )^2 +
\left ( e^{M_{1}} - e^{M_{2}} \right )^2 \right ].
\een
It is evident that on the boundary $q_{(A)} = q_{(B)} = 2 $ and their
first derivatives disappear there. Taking into account the boundary
conditions on $\p \cD$, one can integrate Eq.(\ref{jj}) to obtain the
relation 
\be
\int_{\p \cD} Tr \left ( \cJ_{(i)}^{\dagger} \cJ_{(i)} \right ) = 0.
\label{jj1}
\ee
The expression (\ref{jj1}) implies vanishing of $\cJ_{(A)}$ and
$\cJ_{(B)}$, which in turns causes that $Q_{(i)} = const$ in all region
$\cD$. Because of this fact, $Q_{(i)} = I$ matrix on $\p \cD$, then
$Q_{(i)}= I$ in $\cD$. Therefore we reach to the conclusion that,
$P_{(i) 1} = P_{(i) 2}$ at all points of the region $\cD$ of the
two-dimensional manifold $\cM$.

As was mentioned in Ref.\cite{gs}, the other way of reaching these conclusions
is to observe that, vanishing of (\ref{jj1}) implies the harmonicity of
$q$ in $\cD$ region. Since $q_{(i)} = 2$ on the boundary $\p \cD$, then
it must be equal to the same constant value in the region $\cD$. Thus,
$P_{(i) 1} = P_{(i) 2}$ in $\cD$.

\vspace{0.2cm}
\noindent
The above considerations enables us to formulate 
the main result, the following.

\noindent
{\it Theorem:}
Consider a two-dimensional manifold $\cM$ equipped with a local
coordinates $(r, z)$. Suppose that, $\cD$ is a region in $\cM$ with
boundary $\p \cD$. Let $P_{(i)}$ be hermitian, positive definite
two-dimensional matrices with unit determinants, respectively for 
$i = \tiA, \tiB $ gauge fields. Suppose further that, matrices $P_{(i) 1}$
and $P_{(i)2}$ satisfy Eq.(\ref{dif}), namely
\be
\na \left (  P_{(i) 1}^{-1} \left ( \na Q_{(i)} \right )
P_{(i) 2} \right ) = 0, 
\ee
in the region $\cD$ and have the same boundary conditions on $\p \cD$.
Then, in all points of the region $\cD$, one has that
$P_{(i) 1} = P_{(i) 2}$.
\par
The doubly charged dilaton black holes in $SU(4)$ version of $N = 4, d
= 4$ supergravity are characterized by mass $M$, the $F$-field electric
charge, the $G$-field magnetic charge and the dilaton charge $\Sigma$.
The above theorem envisages that all black holes subject to the same
boundary conditions, as the solution obtained by Kallosh {\it et al.}
in Ref.\cite{kal},
are the same everywhere in the spacetime.

%%%%%%%%%%%%%%%%%%%%%%%%%%%%%%%%%%%%%%%%%%%%%%%%%%%%%%%%%%%%%%%
\subsection{Doubly Charged Black Holes in $SO(4)$ version of $N = 4, d
= 4$ supergravity}
The bosonic part of the $SO(4)$ version of $N = 4$ supergravity in four
dimensions can be described by the action \cite{so,kal}
\be
I_{SO(4)} = \int d^4 x \sqrt{-g} \left [ R - 2(\na \phi)^{2}
- \left ( e^{-2\phi} F_{\alpha \beta} F^{\alpha \beta} +
e^{2 \phi} \tiG_{\ga \delta} \tiG^{\ga \delta} \right )
\right ],\label{so4}
\ee
where the strengthes of the gauge fields are descibed by
$F_{\mu \nu} = 2\na_{[\mu} A_{\nu]}$ and 
$\tiG_{\alpha \beta} = 2\na_{[\alpha} G_{\beta] }$.
The equations derived from the variational principle are as follows:
\ben
R_{\mu \nu} = e^{-2 \phi} \left (
2 F_{\mu \rho} F_{\nu}{}{}^{\rho} - {1 \over 2} g_{\mu \nu}F^2 \right ) +
e^{2 \phi} \left (
2 \tiG_{\mu \rho} \tiG_{\nu}{}{}^{\rho} - 
{1 \over 2} g_{\mu \nu} \tiG^2 \right ) +
2 \na_{\mu} \phi \na_{\nu} \phi, \\
\na_{\mu} \na^{\mu} \phi + {1 \over 2} e^{-2\phi} F^2 -
{1 \over 2} e^{2\phi} \tiG^2 = 0, \\
\na_{\mu} \left ( e^{-2 \phi} F^{\mu \nu} \right ) = 0, \\
\na_{\mu} \left ( e^{2 \phi} \tiG^{\mu \nu} \right ) = 0.
\een
The components of the $U(1)$ gauge fields are in the time 
direction \cite{kal},
namely $F_{\mu \nu} = (A_{0}, 0, 0, 0)$ and $\tiG_{\mu \nu} = (G_{0}, 0
, 0, 0)$. 
As in the proceding paragraph the components of the $U(1)$ gauge fields
are functions depending only on $(r, z)$ coordinates.
The Eqs. of motion in metric (\ref{gij}) satisfy
\be
\na^2 \phi - e^{-2\psi - 2\phi} \left (
A_{0, r}^2 + A_{0, z}^2 \right ) + {e^{- 2\psi + 2 \phi} \over r^2}
\left ( G_{0, r}^2 + G_{0, z}^2 \right ) = 0,
\ee
\be
\na^2 \psi - e^{-2\psi - 2\phi} \left (
A_{0, r}^2 + A_{0, z}^2 \right ) - e^{- 2\psi + 2 \phi}
\left ( G_{0, r}^2 + G_{0, z}^2 \right ) = 0,
\ee
\be
\na^2 A_{0} - 2 \left ( \psi_{, r} + \phi_{, r} \right ) A_{0, r}
- 2 \left ( \psi_{, z} + \phi_{, z} \right ) A_{0, z} = 0,
\ee
\be
\na^2 G_{0} + 2 \left (- \psi_{, r} + \phi_{, r} \right ) G_{0, r}
+ 2 \left ( - \psi_{, z} + \phi_{, z} \right ) G_{0, z} = 0,
\ee
\be
e^{-2\psi - 2\phi} \left (
A_{0, r}^2 - A_{0, z}^2 \right ) + e^{- 2\psi + 2 \phi} 
\left ( G_{0, r}^2 - G_{0, z}^2 \right ) +
\left ( \phi_{,z}^2 - \phi_{, r}^2 \right ) =
\psi_{, r}^2 - \psi_{, z}^2 - {\ga_{, r} \over r},
\ee
\be
{\ga_{, z} \over r} - 2 \psi_{, r} \psi_{, z} =
- 2 e^{-2\psi - 2\phi} A_{0, r} A_{0, z} - 2 e^{ - 2\psi + 2 \phi} 
G_{0, r} G_{0, z} + 2 \phi_{, r} \phi_{, z}.
\ee
Thus, with the substitution
\be
E = - \phi - \psi, \qquad N = \phi - \psi,
\ee
and
\be
\tiA_{0} = {A_{0} \over \sqrt{2}}, \qquad 
\tiG_{0} = {G_{0} \over \sqrt{2}},
\ee
we find that,
Eqs. of motion can be rewritten as follows:
\ben
\label{so1}
\na^2 E + e^{2 E} \na \tiA_{0} \na \tiA_{0} = 0, \\
\na^2 \tiA_{0} + 2 \na E  \na \tiA_{0} = 0, \\
\na^2 N + e^{2 N} \na \tiG_{0}  \na \tiG_{0} = 0, \\
\na^2 \tiG_{0} + 2 \na N \na \tiG_{0} = 0.
\label{so}
\een
Now, the pseudopotentials for gauge fields $\tiA$ and $\tiG$ have 
the same
form as in Eq.(\ref{wa}), where one should substitute the adequate values
of $E, (N)$ and derivatives of the gauge fields under consideration.
Then, using Eq.(\ref{e1}) to introduce the complex scalars, enables to
rewrite the system of Eqs.(\ref{so1}-\ref{so}) as the 
decoupled
two Ernst's like Eqs.
Then, the proof follows the same line as that of the proceding
subsection. Finally one can reach the conclusion, that all points of
the region $\cD$ equipped with the boundary $\p \cD$ on two-dimensional
manifold $\cM$, $P_{(i)1} = P_{(i)2}$. \\
In view of the foregoing uniqueness theorem, we conclude that all
doubly charged
black hole solutions characterized by mass $M$, the $F$-field electric
charge, the $G$-field electric charge and the dilaton charge $\Sigma$,
with the same boundary conditions as found in Ref.\cite{kal} are the same
everywhere in the spacetime.

%%%%%%%%%%%%%%%%%%%%%%%%%%%%%%%%%%%%%%%%%%%%%%%%%%%%%%%%%%%%%%%%%%
\section{Conclusions}
In our work we were studying the doubly charged dilaton black
holes in the bosonic sector of $N = 4, d = 4$ supergravity, being the
low energy limit of the superstring theories. We were interested in the
uniqueness of these solutions. Using the method proposed in Ref.\cite{gu1}
and finding the adequate forms of pseudopotentials and
complex scalars for $SU(4)$ and $SO(4)$ versions of the theory, 
we were able to find that Eqs. of motion could be
arranged in the two Ernst's Eqs. for each of the gauge fields appearing
in the theory. 
These Eqs. give the Ernst's formulation of the generalized Einstein
Eqs. for the bosonic sector of $N = 4, d =4$ supergravity.
Then, using
the idea of G\"urses and Xantopoulous \cite{gx} that Ernst Eqs. are included
in the single matrix Eq., we prove the uniqueness of the previously \cite{kal}
obtained black hole solutions. \\
It will be interesting to find the exact form of the Ernst Eqs. for more
complicated version of the low energy string theory, for instance
Einstein-Maxwell axion-dilaton gravity or $N = 8$ black holes now
intensively studied \cite{ort}. We hope to return to these problems elsewhere.

%%%%%%%%%%%%%%%%%%%%%%%%%%%%%%%%%%%%%%%%%%%%%%%%%%%%%%%%%%%%%%%%%%%%%%%%%%

%%%%%%%%%%%%%%%%%%%%%%%%%%%%%%%%%%%%%%%%%%%%%%%%%%%%%%%%%%%%%%%%%%%

\end{document}